\documentclass[pra,aps,amssymb,twocolumn,noshowpacs]{revtex4-1}
\usepackage{color}
\usepackage{graphicx}
\usepackage{hyperref}
\usepackage{amsmath}
\usepackage{latexsym}
\usepackage{amssymb}
\usepackage{mathrsfs}
\usepackage{mathtools}
\usepackage{textcomp}
\usepackage{layout}
\usepackage{verbatim}
\usepackage{bm}
\usepackage{amsfonts,epsfig}

\begin{document}

\title{Fast atom-photon entangling gates with a superconducting coplanar waveguide}

\date{\today}
\author{Xiao-Feng Shi}
\affiliation{School of Physics, Xidian University, Xi'an 710071, China}

%==============================================================================
\begin{abstract}
  Entanglement between atoms and microwave photons in a superconducting coplanar waveguide~(SCW) can enable hybrid quantum devices and interface static and flying qubits. We study a one-step controlled-Z~(C$_{\text{Z}}$) gate between a neutral atom trapped near a SCW and a microwave mode in the SCW, which is an extension of the gate proposed in [J. D. Pritchard, et.al., Phys. Rev. A 89, 010301(R) (2014)]. The gate protocol is simple and requires one laser pulse for exciting a transition between the ground and Rydberg states of the neutral atom.

\end{abstract}
%==============================================================================
\maketitle

\begin{figure}
\includegraphics[width=2.8in]
{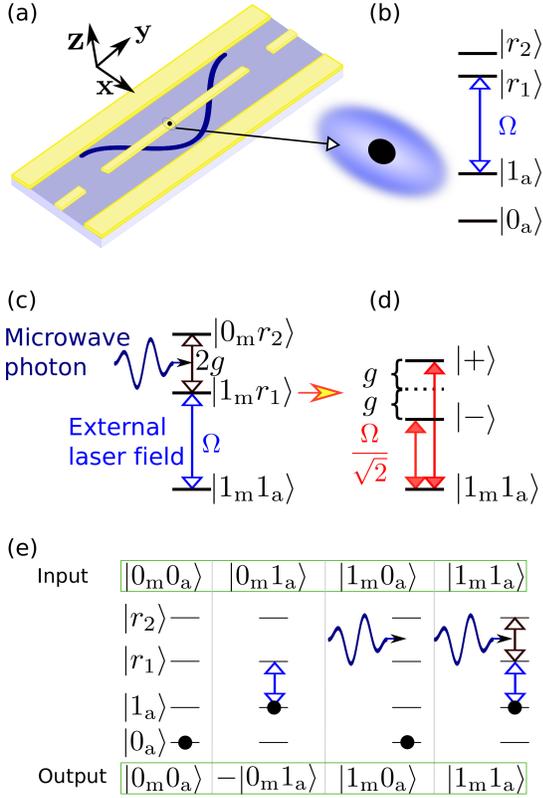}
 \caption{(a) A hybrid quantum system with a SCW resonator and a neutral atom trapped above the electric field antinode of the fundamental mode of the SCW. (b) Four atomic levels of the atom are involved in the model, where $|0_{\text{a}}\rangle$ and $|1_{\text{a}}\rangle$ are two hyperfine-Zeeman ground substates of a $^{133}$Cs atom, and $|r_1\rangle$ and $|r_2\rangle$ are two Rydberg states. (c) $|1_{\text{a}}\rangle$ can be excited to $|r_1\rangle$ by coherent laser fields, and $|r_1\rangle$ and $|r_2\rangle$ can be coupled by the microwave field in the SCW, both in the form of electric dipole transitions. (d) For the input state $|1_{\text{m}}1_{\text{a}}\rangle$ of the system and when the laser field is sent for the transition between $|1_{\text{a}}\rangle$ and $|r_1\rangle$, it is like that $|1_{\text{m}}1_{\text{a}}\rangle$ is excited to $|\pm\rangle\equiv (|1_{\text{m}} r_1\rangle\pm|0_{\text{m}} r_2\rangle)/\sqrt{2}$ with a detuning $\pm g$, where $2g$ is the single-photon Rabi frequency of the atom-microwave field coupling. (e) An atom-photon C$_{\text{Z}}$ gate is realized by a $2\pi$ pulse of laser excitation.  \label{fig001}} 
\end{figure}

\section{introduction}
Scalable quantum computing is capable to solve problems beyond the computational power of classical computers~\cite{Shor1997,Grover1997,Grover1998}, for which several promising candidates are being explored to realize quantum computers on the physical level~\cite{Ladd2010}. Recently, neutral atom arrays emerge as a new promising candidate for this purpose because they can have long coherence times on the order of second~\cite{Wang2016} or even tens of seconds~\cite{Young2020}, can support over two hundred individual qubits in one quantum register~\cite{Schymik2020,Semeghini2021,Ebadi2021}, and can enable both one~\cite{Olmschenk2010,Xia2015,Wang2016} and two-qubit~\cite{Madjarov2020} logic operations with fidelities near the threshold of fault-tolerant quantum computing~\cite{Crow2016}.

On the other hand, quantum communication usually relies on flying qubits in the form of microwave or optical photons that travel in either free space~\cite{Vazquez-Castro2021} or guided pathways like optical fibers or waveguides~\cite{PhysRevLett.126.250502}. Long-distance quantum communication was usually studied with optical photons~\cite{pan2012}, and there are also ongoing efforts to explore short- to medium-distance quantum communication by microwave photons both theoretically~\cite{Candia2021} and experimentally~\cite{Barzanjeh2020,Fedorov2021} partly because of the mature study on circuit quantum electrodynamics~\cite{Kokkoniemi2020,RevModPhys.93.025005} and several advantages~(such as in biomedical use) possessed by the low power feature of microwave signals~\cite{Karsa2020}. Therefore, it is useful to enable a quantum interface between static qubits and flying microwave qubits so that the quantum information stored in long-lived quantum registers can be transferred to photons so as to be further sent to a distant destination.

In this paper, we study a controlled-Z~(C$_{\text{Z}}$) gate to rapidly entangle microwave photons in a superconducting coplanar waveguide~(SCW) and a nearby neutral atom. The proposed C$_{\text{Z}}$ gate between the atomic qubit and the microwave mode is of identical form as the one proposed in Ref.~\cite{Pritchard2014}, but is relatively simple: it requires one laser pulse to excite the atomic state between ground and Rydberg states. In comparison, the C$_{\text{Z}}$ gate in Ref.~\cite{Pritchard2014} needs several steps: a laser pulse for Rydberg excitation, two switches of the cavity~(or atomic transition) frequency, and another laser pulse for Rydberg deexcitation.

\section{A one-step atom-photon C$_{\text{Z}}$ gate}
Following Ref.~\cite{Pritchard2014}, we assume that there is at most one microwave photon in the SCW resonator and use $|0_{\text{m}}\rangle$ and $|1_{\text{m}}\rangle$ to label the microwave state in the SCW with zero and one photon, respectively. We use $|0_{\text{a}}\rangle$ and $|1_{\text{a}}\rangle$ to label the two hyperfine-Zeeman ground substates of a heavy alkali-metal atom such as $^{133}$Cs. The atom is trapped above the SCW and is near the antinode of the microwave field as schematically shown in Fig.~\ref{fig001}(a). The hybrid atom-photon C$_{\text{Z}}$ gate realizes the state mapping
\begin{eqnarray}
  &&\{|0_{\text{m}}0_{\text{a}}\rangle,|0_{\text{m}}1_{\text{a}}\rangle,|1_{\text{m}}0_{\text{a}}\rangle,|1_{\text{m}}1_{\text{a}}\rangle\}\nonumber\\
  &&\rightarrow \{|0_{\text{m}}0_{\text{a}}\rangle,-|0_{\text{m}}1_{\text{a}}\rangle,|1_{\text{m}}0_{\text{a}}\rangle,|1_{\text{m}}1_{\text{a}}\rangle\}, \label{gate01}\nonumber
\end{eqnarray}
and is implemented by one laser pulse to excite the transition between $|1_{\text{a}}\rangle$ and a Rydberg state $|r_1\rangle$ with a Rabi frequency $\Omega$ and duration $2\pi/\Omega$. When $\Omega$ is much smaller than the frequency separation between $|0_{\text{a}}\rangle$ and $|1_{\text{a}}\rangle$, the other state $|0_{\text{a}}\rangle$ of the atomic qubit is not excited, so that $|0_{\text{m}}0_{\text{a}}\rangle$ stays intact, while $|0_{\text{m}}1_{\text{a}}\rangle$ acquires a $\pi$ phase after a full Rabi cycle $|0_{\text{m}}1_{\text{a}}\rangle\rightarrow -i|0_{\text{m}}r_1\rangle\rightarrow-|0_{\text{m}}1_{\text{a}}\rangle$. On the other hand, the microwave photon is far detuned from the magnetic dipole transition between $|0_{\text{a}}\rangle$ and $|1_{\text{a}}\rangle$, and meanwhile this magnetic dipole coupling is extremely weak~\cite{Hafezi2012}, so that $|1_{\text{m}}0_{\text{a}}\rangle$ is not excited and stays intact. Below, we show that there is a condition for $|1_{\text{m}}1_{\text{a}}\rangle$ to acquire no phase change so as to realize the C$_{\text{Z}}$ gate. 

The state $|1_{\text{m}}1_{\text{a}}\rangle$ is coupled to $|1_{\text{m}} r_1\rangle$ by the laser field, but $|1_{\text{m}} r_1\rangle$ is further coupled to another state because we consider that the microwave photon is resonant with the transition between $|r_1\rangle$ and a nearby Rydberg state $|r_2\rangle$, i.e., the microwave frequency is equal to the frequency difference between $|r_1\rangle$ and $|r_2\rangle$. With rotating-wave approximation in a rotating frame, the atom-microwave field coupling is described by the Hamiltonian 
\begin{eqnarray}
  \hat{H}_{\text{m-a}} &=&\hbar g |0_{\text{m}} r_2\rangle\langle 1_{\text{m}} r_1|+ \text{H.c.},
\label{microwave-atom01}
\end{eqnarray}
where $\hbar$ is the reduced Planck constant, and $2g$ is the single-photon Rabi frequency which is assumed to be a real variable for brevity. The above Hamiltonian can be diagonalized as
\begin{eqnarray}
  \hat{H}_{\text{m-a}} &=&\hbar g (|+\rangle\langle +|- |-\rangle\langle -|),
\label{microwave-atom02}\nonumber
\end{eqnarray}
where 
\begin{eqnarray}
|\pm\rangle &=& (|1_{\text{m}} r_1\rangle\pm|0_{\text{m}} r_2\rangle)/\sqrt{2}. \nonumber
\end{eqnarray}
In the basis of $|+\rangle,|-\rangle$, and $|1_{\text{m}}1_{\text{a}}\rangle$, the Hamiltonian of atom-laser-microwave field interaction is
\begin{eqnarray}
  \hat{H}_{11}&=&\frac{ \hbar\Omega}{2\sqrt{2}} \left(
  \begin{array}{ccc}
    2\eta&0&1\\
    0&- 2\eta& 1 \\
   1 & 1 &0
    \end{array}
  \right),\label{hamiltonian11input}
\end{eqnarray}
where the subscript in $\hat{H}_{11}$ denotes that it is applicable for the input state $|1_{\text{m}}1_{\text{a}}\rangle$, and
\begin{eqnarray}
  \eta &=& \sqrt{2}g/\Omega.\nonumber
\end{eqnarray}
By exploring the time evolution operator and the eigenvalues of Eq.~(\ref{hamiltonian11input}), one can show that when $|\eta|=\sqrt{6}/2$ is satisfied and when we start from an initial wave function $|\psi(0)\rangle$ superposed of $|+\rangle,|-\rangle$, and $|1_{\text{m}}1_{\text{a}}\rangle$, the wavefunction is exactly restored after one $2\pi$ pulse, i.e., $|\psi(t)\rangle=|\psi(0)\rangle$ with $t=2\pi/\Omega$; more details can be found in Appendix~\ref{appendixA}. For our purpose, this means that the input state $|1_{\text{m}}1_{\text{a}}\rangle$ stays intact upon the completion of the $2\pi$ pulse of laser excitation in the condition $|\eta|=\sqrt{6}/2$.

\begin{figure}
\includegraphics[width=3.3in]
{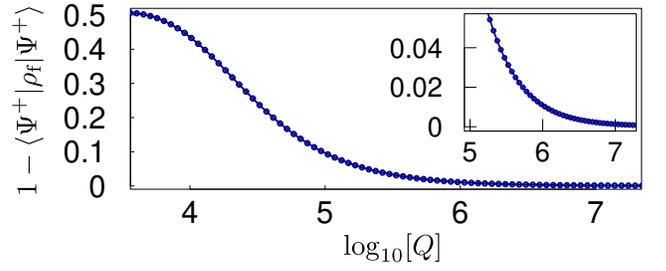}
 \caption{Infidelity of Bell-state preparation by using the C$_{\text{Z}}$ gate as a function of Q with $2g/\sqrt{3}=\Omega=2\pi\times1$~MHz, $T=50$~mK, $\omega_c=2\pi\times5.037$~GHz, $(\tau_1,\tau_2)=(0.82,~1.97)~$ms, and $\hat{a}$ and $\hat{a}^\dag$ are expanded in Fock space with photon number up to 5. Simulation with Eq.~(\ref{Lindblad1}) was via QuTiP~\cite{Johansson2012,Johansson2013} except that the fidelity $\mathcal{F}=\langle \Psi^+|\rho_{\text{f}}|\Psi^+\rangle$ here is square of the `fidelity' function of QuTiP or~\cite{Pritchard2014}. The Bell-state fidelity is $94.9\%$ when $Q=2\times10^5$, and increases to $98.9\%$ when $Q=10^6$. The inset shows data around $Q=10^6$. \label{fig-infidQ}} 
\end{figure}

\begin{figure}
\includegraphics[width=3.2in]
{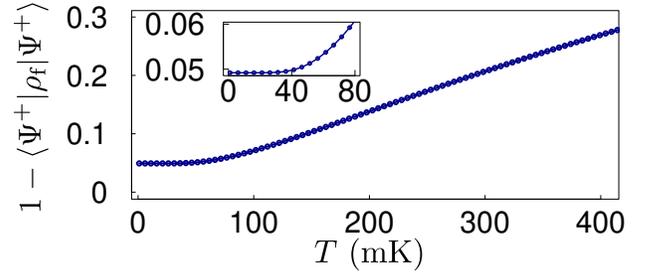}
 \caption{ Infidelity of preparation of the Bell state as a function of T when $Q=2\times10^5$; other parameters are the same to those in Fig.~\ref{fig-infidQ}. The inset shows infidelities around $T=50$~mK. \label{fig-infidT}} 
\end{figure}

\section{Operation fidelity}\label{sec03}
The fidelity of the C$_{\text{Z}}$ gate is subject to the incoherent coupling between the atom and the thermal microwave photons in the environment, the decay of the microwave and atomic qubits, and the fluctuation of atom-photon coupling due to the fluctuation of the atomic position. To investigate their influence on the gate, we study the preparation fidelity $\mathcal{F}=\langle \Psi^+|\rho_{\text{f}}|\Psi^+\rangle$ of atom-photon Bell state $|\Psi^+\rangle = (|0_{\text{m}}1_{\text{a}}\rangle+|1_{\text{m}}0_{\text{a}}\rangle)/\sqrt{2}$. Here, $\rho_{\text{f}}$ is the density matrix of the state prepared from the initial product state $(|0_{\text{m}}0_{\text{a}}\rangle+|1_{\text{m}}0_{\text{a}}\rangle)/\sqrt{2}$ by $H_{\text{a}}$C$_{\text{Z}}H_{\text{a}}$, where $H_{\text{a}} = (|x_{\text{m}}0_{\text{a}}\rangle+ |x_{\text{m}}1_{\text{a}}\rangle)\langle x_{\text{m}}0_{\text{a}}|/\sqrt{2} + (|x_{\text{m}}0_{\text{a}}\rangle- |x_{\text{m}}1_{\text{a}}\rangle)\langle x_{\text{m}}1_{\text{a}}|/\sqrt{2}$ is the Hadamard gate for the atomic state and $x_{\text{m}}=0$ or $1$. To focus on the operation accuracy of the C$_{\text{Z}}$ gate, we assume that the two Hadamard gates operate ideally, so that the Bell-state fidelity reflects the fidelity of the C$_{\text{Z}}$ gate.

The fidelity can be studied with a setup analyzed in Ref.~\cite{Pritchard2014}, where a cesium atom is trapped above a SCW with $Q=2\times10^5$ and $\omega_c=2\pi\times5.037$~GHz. This $\omega_c$ corresponds to the transition between $|r_{1}\rangle=|90S_{1/2},m_J=1/2\rangle$ and $|r_{2}\rangle=|90P_{3/2},m_J=1/2\rangle$, for which the Rydberg-state lifetimes are $(\tau_1,\tau_2)=(0.82,~1.97)~$ms~\cite{Beterov2009} below $400$~mK. The dynamics can be studied by the optical Bloch equation in the Lindblad form,
\begin{eqnarray}
 \frac{d\rho}{dt} &=& i (\rho\hat{H}- \hat{H} \rho)/\hbar +\sum_{i=0}^3\left[2\hat{c}_i\rho\hat{c}_i^\dag- \hat{c}_i^\dag\hat{c}_i \rho-\rho\hat{c}_i^\dag\hat{c}_i  \right]/2,\nonumber\\
 \label{Lindblad1}
\end{eqnarray}
where the Hamiltonian $\hat{H}$ and collapse operators $\hat{c}_i$ are given by 
\begin{eqnarray}
 \hat{H} &=& \hbar( g|r_2\rangle\langle r_1|\hat{a}+ \Omega|r_1\rangle\langle1|/2) + \text{H.c.},\nonumber\\
 \hat{c}_0 &=& |g\rangle\langle r_1|/\sqrt{\tau_1},\nonumber\\
 \hat{c}_1 &=& |g\rangle\langle r_2|/\sqrt{\tau_2} ,\nonumber\\
 \hat{c}_2 &=& \hat{a}\sqrt{(\overline{n}_{\text{th}}+1)\omega_c/Q}  ,\nonumber\\
 \hat{c}_3 &=& \hat{a}^\dag\sqrt{\overline{n}_{\text{th}}\omega_c/Q}.\nonumber  
\end{eqnarray}
where $|g\rangle$ is a virtual reservoir state in the ground-state manifold that is assumed dark to the external fields, $\hat{a}$ is the annihilation operator of the SCW microwave mode, and $\overline{n}_{\text{th}}=\left[\text{exp}(\hbar\omega_c/k_{\text{B}}T)-1 \right]^{-1}$ gives the mean number of thermal microwave photons of the SCW mode. 

One major influence over the gate fidelity is from the coupling between the cavity mode and the thermal photons in the environment. Two factors determine the significance of this coupling, the quality factor of the cavity $Q$, and the temperature $T$. The increase of the Bell-state fidelity via the increase of $Q$ is shown in Fig.~\ref{fig-infidQ} with $T=50$~mK. When $Q$ is fixed at $2\times10^5$, the decrease of fidelity when $T$ increases is shown in Fig.~\ref{fig-infidT}. The data in Figs.~\ref{fig-infidQ} and~\ref{fig-infidT} are calculated with $g/2\pi\approx0.866$~MHz. If $g$ is larger, then the fidelities increase because the incoherent processes have a less influence thanks to the faster gate speed. For example, $F$ is $95.0\%$ at $T=40$~mK in Fig.~\ref{fig-infidT}, but it increases to $97.8\%$ if $g/2\pi\approx2$~MHz at the same temperature, which shows that the fidelity here is comparable to that of Ref.~\cite{Pritchard2014}; when comparing to Ref.~\cite{Pritchard2014}, we shall note that here we did not consider the preparation of the initial product state, and there is difference in defining the fidelity, i.e., $\langle \Psi^+|\rho_{\text{f}}|\Psi^+\rangle$ defines $\mathcal{F}$ here, but its square root is the definition of fidelity in Ref.~\cite{Pritchard2014}.

Another noise is from the fluctuation of $g$ due to the position fluctuation of atom. We take the field distribution of the SCW analyzed in Ref.~\cite{Pritchard2014} as an example, where the schematic of the SCW is shown in Fig.~\ref{fig001}(a). For brevity, we consider trapping the atom above the center of the center conductor by a laser light sent along $\mathbf{y}$, and the trap is switched off when the atom-photon gate is in process. In this case, Ref.~\cite{Pritchard2014} showed that the atom shall be trapped at a height $z\approx17~\mu$m if we want $g/2\pi=\sqrt{3}/2$~MHz in our case, and $g/2\pi$ varies almost linearly in $z$ with a slope about $\varsigma=0.12$~MHz$/\mu$m. The strength of the microwave field around the SCW can be numerically studied~\cite{Hattermann2017}, which showed that the field above the central line of the centre conductor is relatively smooth along a direction parallel to the $\mathbf{x}\mathbf{y}$ plane. So, we can ignore the fluctuation of $g$ due to the position fluctuation of the atom in the $\mathbf{x}\mathbf{y}$ plane, but only consider the fluctuation of the atomic position along $\mathbf{z}$. By using Eq.~(\ref{Lindblad1}) for simulating the fidelity $\mathcal{F}(g+\varsigma z)$ when the atom is vertically away from the trap center by $z$,  the average Bell-state fidelity can be approximated by 
\begin{eqnarray}
\overline{\mathcal{F}} &=& \frac{1}{\sqrt{2\pi}\sigma}\int e^{-\frac{z^2}{2\sigma^2}}\mathcal{F}(g+\varsigma z) dz,\label{infidZ}
\end{eqnarray}
when the value of $\Omega$ is fixed at $2g/\sqrt{3}$. The approximation for a fixed $\Omega$ is based on that the fluctuation of $\Omega$ can be made negligible by focusing the Rydberg laser fields with a large beam waist and Rayleigh length and sending them along a direction near to the travelling direction of the tweezer field. For example, excitation of high-lying Rydberg states was realized with fields of beam waist about $10~\mu$m in Ref.~\cite{Isenhower2010}, but the fields for trapping atoms can be more focused, e.g., the r.m.s. position fluctuation of the atom perpendicular to the beam direction was $\sigma=0.27~\mu$m in the experiment reported in Ref.~\cite{Graham2019}. 

\begin{figure}
\includegraphics[width=3.2in]
{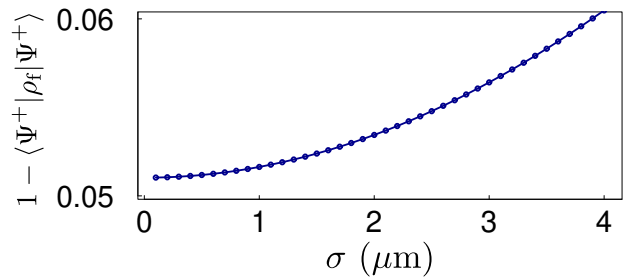}
 \caption{ Infidelity of preparation of the Bell state as a function of r.m.s. position fluctuation of the atom which changes the atom-photon coupling from $g$ to $g+\varsigma z$, where $g/2\pi=\sqrt{3}/2$~MHz, $\varsigma/2\pi=0.12$~MHz$/\mu$m, and $z$ is normally distributed. The fidelity was averaged via Eq.~(\ref{infidZ}) in which each $\mathcal{F}(g+\varsigma z)$ was calculated by using Eq.~(\ref{Lindblad1}) with $Q=2\times10^5$; other parameters are the same to those in Fig.~\ref{fig-infidQ}.  \label{fig-infidZ}} 
\end{figure}

The results of numerical simulation given in Fig.~\ref{fig-infidZ} show that when $g$ changes due to the position fluctuation of the atom, the Bell-state fidelity does not decrease so quickly as the cases in Figs.~\ref{fig-infidQ} and~\ref{fig-infidT}. The reason is that the change of $g$ only influences the state component $|1_{\text{m}}1_{\text{a}}\rangle$ in the transition chain $|1_{\text{m}}1_{\text{a}}\rangle\rightarrow |1_{\text{m}} r_1\rangle\rightarrow|0_{\text{m}} r_2\rangle$, or more accurately, in the transition from $|1_{\text{m}} r_1\rangle$ to $|0_{\text{m}} r_2\rangle$. Importantly, our gate protocol only involves partial population in $|1_{\text{m}} r_1\rangle$, so that the fluctuation of $g$ does not impart much effect. In contrast, other gate protocols involving full population of $|1_{\text{m}} r_1\rangle$ can have a larger gate error due to the fluctuation of $g$ as discussed in Appendix~\ref{appendixA}.

\section{Discussion}
The C$_{\text{Z}}$ gate of this paper imposes a state transform identical to that of Ref.~\cite{Pritchard2014}, but the two gate protocols are different.

First, the gate in this paper needs one laser pulse, while the one in~\cite{Pritchard2014} needs several steps: (i) Use a $\pi$ pulse of laser field to drive the transition between the atomic qubit state and the Rydberg state $|r_1\rangle$ when the the transition $|r_1\rangle\leftrightarrow|r_2\rangle$ is highly offresonant with the microwave mode. (ii) Tune the transition $|r_1\rangle\leftrightarrow|r_2\rangle$ to be resonant with the microwave mode, wait for a time $\pi/g$, and restore the highly off-resonant condition again. (iii) repeat step (i). The protocol in~\cite{Pritchard2014} hinges on the capability to rapidly tune the detuning between the transition frequency for $|r_1\rangle\leftrightarrow|r_2\rangle$ and the frequency of the microwave mode, which may be possible either by using Stark shift in the Rydberg states, or by using intracavity superconducting quantum interference devices~\cite{Sandberg2008,Wang2013}. 

Second, the total gate duration in this paper is $2\pi/\Omega=\sqrt{3}\pi/g$, while that in~\cite{Pritchard2014} is $2\pi/\Omega_{\text{o}}+\pi/g_{\text{o}}$; here the subscript o is to distinguish parameters in different protocols. This means that the gate speed in this paper is faster by $\pi/g_{\text{o}}$ if $\Omega=\Omega_{\text{o}}$, or differ from that of~\cite{Pritchard2014} by $2\pi/\Omega_{\text{o}}-(\sqrt{3}-1)\pi/g_{\text{o}}$ if $g=g_{\text{o}}$. This latter condition means that the gate in~\cite{Pritchard2014} is faster if $\Omega_{\text{o}}/g_{\text{o}}>\sqrt{3}+1$ when $g=g_{\text{o}}$. The gate in this paper needs the condition $g=\sqrt{3}\Omega/2$, but the gate of~\cite{Pritchard2014} does not depend on any relation between $\Omega_{\text{o}}$ and $g_{\text{o}}$, which means that the gate in Ref.~\cite{Pritchard2014} can also be very fast if $\Omega_{\text{o}}$ can be large. Ref.~\cite{Pritchard2014} considered a setup where the cesium atom is trapped at a point about $10~\mu$m above a SCW with a quality factor $Q=2\times10^5$ and angular frequency $\omega_c=2\pi\times5.037$~GHz of the microwave field mode, and found that it is possible to achieve $g=2\pi\times2$~MHz with $\omega_c$ corresponding to the transition between the Rydberg states $|r_{1}\rangle=|90S_{1/2},m_J=1/2\rangle$ and $|r_{2}\rangle=|90P_{3/2},m_J=1/2\rangle$. Detailed analyses in Ref.~\cite{Pritchard2014} showed that in order to minimize gate errors from atomic and photonic incoherent dynamics, choosing Rydberg states around principal quantum number $n=90$ is useful. But the excitation of high-lying Rydberg states with $n>90$ involves small dipole matrix elements, so that the value of $\Omega$ is usually smaller than $2\pi\times1$~MHz~\cite{Isenhower2010,Zhang2010} with reasonable laser powers. However, the gate in Ref.~\cite{Pritchard2014} can be in principle fast when very large Rabi frequency is realizable.

Third, as discussed at the end of Sec.~\ref{sec03} and in Appendix~\ref{appendixA}, the gate in this paper is less vulnerable to the fluctuation of $g$ caused by, e.g., the position randomness of the atom. 

The main advantage of the gate in this paper is that it is simpler concerning its implementation. Though it can be fast, the fidelity concerning a practical experimental realization is mainly determined by the incoherent processes from the background thermal photons in the SCW; the numerical simulation shown in Sec.~\ref{sec03} indicates that the fidelity here is similar to that of Ref.~\cite{Pritchard2014} with similar aotm-photon coupling rates and incoherent processes~(we note, however, that $\Omega_{\text{o}}$ used in Ref.~\cite{Pritchard2014} had a large value, $2\pi\times10$~MHz).

%==============================================================================
%=========conclusion
\section{Conclusion}\label{sec04}
We study a C$_{\text{Z}}$ gate between a neutral atom trapped near a SCW and a microwave mode in the SCW. The gate is realized by a $2\pi$ pulse of laser fields for the excitation of a ground-Rydberg transition, which extends the gate proposed in Ref.~\cite{Pritchard2014} to a simpler version. The fast and simple entanglement generation between atoms and microwave photons in a SCW can simplify operations of atom-photon hybrid quantum devices and bring opportunities to interface static and flying qubits.

%=========conclusion-End
%==============================================================================

\section*{ACKNOWLEDGMENTS}
The author thanks T. A. B. Kennedy and Yan Lu for useful discussions. This work is supported by the National Natural Science Foundation of China under Grants No. 12074300 and No. 11805146, the Natural Science Basic Research plan in Shaanxi Province of China under Grant No. 2020JM-189, and the Fundamental Research Funds for the Central Universities.

\begin{appendix}
\section{Time evolution for the input state $|1_{\text{m}}1_{\text{a}}\rangle$}\label{appendixA}
Here, we show the time evolution of the wavefunction if the input state is $|1_{\text{m}}1_{\text{a}}\rangle$ for the C$_{\text{Z}}$ gate when ignoring the incoherent processes. Equation~(\ref{hamiltonian11input}) can be diagonalized by its three eigenvectors
\begin{eqnarray}
  |\epsilon_+\rangle &=&[(\eta+\sqrt{0.5+\eta^2})|+\rangle-(\eta-\sqrt{0.5+\eta^2})|-\rangle\nonumber\\
    &&+|1_{\text{m}}1_{\text{a}}\rangle]/\mathcal{N}(\eta),\nonumber\\
  |\epsilon_-\rangle &=&  [(\eta-\sqrt{0.5+\eta^2})|+\rangle-(\eta+\sqrt{0.5+\eta^2})|-\rangle\nonumber\\
    &&+|1_{\text{m}}1_{\text{a}}\rangle]/\mathcal{N}(\eta),\nonumber\\
  |\epsilon_0\rangle &=& [-|+\rangle+|-\rangle+2\eta|1_{\text{m}}1_{\text{a}}\rangle]/\mathcal{N}(\eta),\nonumber
\end{eqnarray}
with their three eigenvalues
\begin{eqnarray}
  \epsilon_\pm &=&\pm\hbar\Omega \mathcal{N}(\eta)/(2\sqrt{2}),\nonumber\\
  \epsilon_0 &=& 0,\nonumber
\end{eqnarray}
where the $\eta$-dependent normalization factor is
\begin{eqnarray}
  \mathcal{N}(\eta)&=& \sqrt{2+4\eta^2}.\nonumber
\end{eqnarray}
The above eigensystem leads to
\begin{eqnarray}
|1_{\text{m}}1_{\text{a}}\rangle  &=& [ |\epsilon_+\rangle+ |\epsilon_-\rangle+2\eta |\epsilon_0\rangle  ]/\mathcal{N}(\eta),\nonumber
\end{eqnarray}
which means that starting from the initial state $|\psi(0)\rangle=|1_{\text{m}}1_{\text{a}}\rangle$, the wavefunction at $t$ is
\begin{eqnarray}
  |\psi(t)\rangle  &=& [ e^{-it \epsilon_+ /\hbar}|\epsilon_+\rangle+ e^{-it \epsilon_-/\hbar }|\epsilon_-\rangle+2\eta e^{-it \epsilon_0/\hbar} |\epsilon_0\rangle  ]/\mathcal{N}(\eta)\nonumber\\
  &=& [ e^{-it \epsilon_+ /\hbar}|\epsilon_+\rangle+ e^{it \epsilon_+/\hbar}|\epsilon_-\rangle+2\eta  |\epsilon_0\rangle  ]/\mathcal{N}(\eta)\nonumber.
\end{eqnarray}
So, when we have the condition $|\eta|=\sqrt{6}/2$, the wavefunction becomes $|\psi(t)\rangle=|1_{\text{m}}1_{\text{a}}\rangle$ at the moment $t=2\pi/\Omega$ because it leads to $t \epsilon_+ =2\pi$, i.e., a $2\pi$ pulse is sufficient to complete the gate. From the above equations, we can see that for a general superposition of $|1_{\text{m}}1_{\text{a}}\rangle, |1_{\text{m}} r_1\rangle$, and $|0_{\text{m}} r_2\rangle$, it is completely restored to the initial state upon the application of a $2\pi$ pulse.

\begin{figure}
\includegraphics[width=3.2in]
{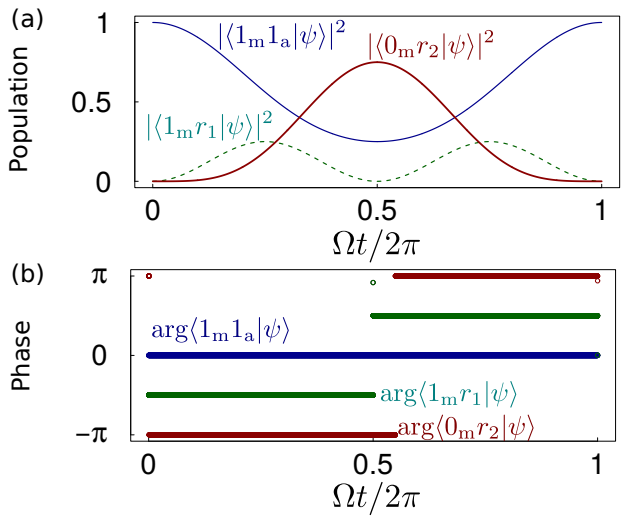}
 \caption{Time evolution of the input state $|1_{\text{m}} 1_{\text{a}}\rangle$ in the gate. Simulation was performed by using Eq.~(\ref{microwave-atom02}) with incoherent processes ignored; detailed analyses on decoherence were given in the main text. (a) and (b) use blue, green, and red symbols to show the population and phase of the components $|1_{\text{m}}1_{\text{a}}\rangle$, $|1_{\text{m}} r_1\rangle$, and $|0_{\text{m}} r_2\rangle$ in the wavefunction $|\psi\rangle$, respectively.  \label{fig002}} 
\end{figure}

The time evolution for the input state $|1_{\text{m}}1_{\text{a}}\rangle$ can be shown by numerical simulation with the original Hamiltonian,
\begin{eqnarray}
  \hat{H}_{\text{m-a}} &=&\text{Eq}.~(\ref{microwave-atom01})+( \hbar \Omega |1_{\text{m}} r_1\rangle\langle 1_{\text{m}} 1_{\text{a}}|+ \text{H.c.})/2.
  \label{microwave-atom02}
\end{eqnarray}
 The result of simulation is shown in Fig.~\ref{fig002} with the condition $\eta= \sqrt{2}g/\Omega=\sqrt{6}/2$, which exactly matches with the analytical results. Of course, when decay of the Rydberg states and cavity mode is included, and when the incoherent processes due to the thermal microwave photons is included, there will be errors as analyzed in the main text.

\begin{figure}
\includegraphics[width=3.2in]
{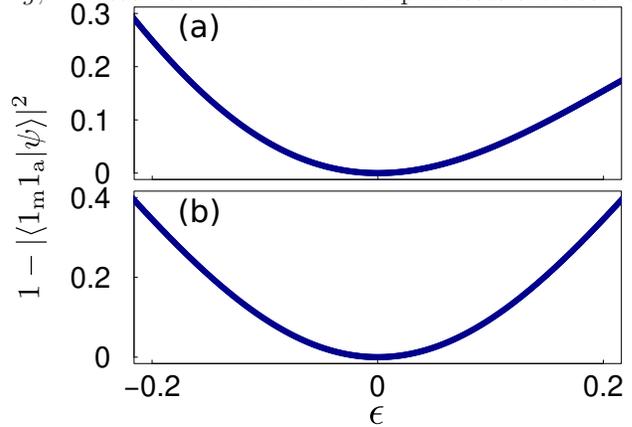}
 \caption{Population error in $|1_{\text{m}} 1_{\text{a}}\rangle$~(when it is the initial wavefunction) as a function of $\epsilon$ with atom-photon coupling rate $g(1+\epsilon)$, where $g$ is the desired value of atom-photon coupling rate. (a) shows results by using the gate sequence of this paper, and (b) shows results from the gate protocol of Ref.~\cite{Pritchard2014}. Incoherent processes are ignored, and the value of $\Omega$ is assumed to be equal to the desired value so as to show the effect only from the fluctuation of $g$.   \label{fig-Comparison}} 
\end{figure}

 A special property for the gate protocol here is that the gate fidelity is not so sensitive to the change of $g$ as the previous method. To show this, we note that among the four input eigenstates, only $|1_{\text{m}}1_{\text{a}}\rangle$ is subject to the atom-photon coupling with fluctuating $g$. In order to investigate only the gate robustness against the fluctuation of $g$, we assume all the incoherent processes are absent, and only study if we start from $|1_{\text{m}}1_{\text{a}}\rangle$, how much population remains there at the end of the gate sequence when $g(1+\epsilon)$, instead of the desired $g$, is used as the atom-photon coupling in the gate, where $\epsilon$ is the relative error in the coupling rate. Shown in Fig.~\ref{fig-Comparison}(a) is the population error in $|1_{\text{m}}1_{\text{a}}\rangle$ as a function of $\epsilon$. The errors are asymmetric with respect to $\epsilon$ due to that the population transfer is controlled not only by $g(1+\epsilon)$ but also by $\Omega$, and are $0.25$ and $0.155$ for $\epsilon=-0.2$ and $0.2$, respectively. If the gate protocol from Ref.~\cite{Pritchard2014} is used, then the population error is shown in Fig.~\ref{fig-Comparison}(b), and is equal to $0.35$ for both $\epsilon=-0.2$ and $0.2$. It is interesting to note that the curve in Fig.~\ref{fig-Comparison}(a) is asymmetric, while that in Fig.~\ref{fig-Comparison}(b) is symmetric with respect to $\epsilon$. To understand the difference in Fig.~\ref{fig-Comparison}(a) and Fig.~\ref{fig-Comparison}(b), we note that for the method in this paper, the population transfers from $|1_{\text{m}}1_{\text{a}}\rangle$ to $|1_{\text{m}} r_1\rangle$ only partially, which further transfers to $|0_{\text{m}} r_2\rangle$ and back via the atom-photon coupling. But for the method in Ref.~\cite{Pritchard2014}, the population is supposed to completely transfer to $|1_{\text{m}} r_1\rangle$ first, and then completely to $|0_{\text{m}} r_2\rangle$ and back, which is why it has a larger error.
 
\end{appendix}

%===============================================================================

%

%===============================================================================
\end{document}